# Difficulties of Model Construction in Optics

# Probleme der Modellbildung in der Optik

Martin Erik Horn, Antje Leisner, Helmut F. Mikelskis

University of Potsdam, Physics Education Research Group,
Am Neuen Palais 10, D - 14469 Potsdam, Germany

E-Mail:
marhorn@rz.uni-potsdam.de
aleisner@rz.uni-potsdam.de
mikelskis@rz.uni-potsdam.de

**Abstract**

Throughout all years of study, students of physics are confronted with the question "what exactly is light": a question that is impossible to answer correctly and, therefore, continuously discussed within the framework of models. Numerous models have been introduced for this purpose and subsequently compete with one another for validity. On the one hand, this may lead to learning difficulties. Yet, on the other hand, this may be considered a didactic chance, if thinking about the model is compounded with the consideration of models in general.
Research in secondary school level II physics classes and with physics students provides insight into the processes of model construction as well as that of overcoming models, which will be presented and analyzed here. The possibilities which computer simulations provide for the discussion of various models of light are demonstrated and their influence on a structured construction of models will be discussed.

**Kurzfassung**

Durch alle Klassenstufen hindurch werden die Schülerinnen und Schüler im Physikunterricht mit der Frage konfrontiert, was Licht eigentlich sei - eine Frage, die korrekt nicht zu beantworten ist und deshalb durchgehend unter Rückgriff auf Modelle erörtert wird. Dabei treten zahlreiche Modellvorstellungen in Konkurrenz miteinander, was einerseits zu Lernschwierigkeiten führt, andererseits jedoch auch als didaktische Chance begriffen werden kann, wenn neben das Denken in Modellen ein Denken über Modelle tritt.
Anhand von Untersuchungen im Physikunterricht der Sekundarstufe II und mit Physikstudenten werden diese Modellbildungs- und Modellüberwindungsprozesse dargestellt und analysiert. Die Chancen und Möglichkeiten, die ein Einsatz vom Computersimulationsprogrammen bei einer Diskussion der unterschiedlichen Lichtmodelle im Unterricht bietet, werden aufgezeigt und ihr Einfluss auf eine strukturierte Modellbildung erörtert.

## 1. Introduction

Light, in its complex physical nature, has been a great puzzle to mankind for ages, and presents many difficulties in our attempts to understand it. As Einstein once said, "I would like to spend the rest of my life thinking about what light is."

In the course of time, therefore, varying models of light have been developed, which help us to be able to clarify light and its behavior: the ray model, the wave model according to Huygens, the particle model according to Newton, the photon model and the phasor model according to Feynman. The models co-exist and are all "equally correct." A model can be neither right nor wrong, but rather useful or useless, since it serves the purpose of perspective and presentation which lay outside of human perception. One can therefore only determine if a model is appropriate for a certain phenomenon which, for the students, means that it can be explained. In this sense, different people can explain one phenomenon with different models, as long as the approach is logical and conclusive.

At the moment the ray model is mainly taught in the beginning physics courses, while the wave model is first introduced in level I classes and the photon model in level II classes at the *Gymnasium*. Newton's light-particle model and Feynman's phasor model hardly find mention, although there are already experiments and suggestions for teaching the phasor model to level II classes [1]. The particle model is only introduced as an aside to the handling of the photon model, when it is necessary to convey that this is not sufficient in clarifying interferences or the photo effect.

The five models of light can be classified into different categories of models (see figure 1). They are differentiated according to concrete and abstract types of models [2]. The former are also designated as models of objects and include, for example, the globe and model train. The abstract model are subdivided into iconic (pictorial) models and into symbolic or mathematical abstract models [3]. Iconic models result from the human spirit, imagination, which man creates from something real but otherwise intangible. According to the extent of the relationship to reality, one can divide the iconic models into two forms: those, that develop from the idealization of reality and those, that only exist in the imagination – hardly having any relation to reality.

The ray model of light is an idealization, while the particle and wave models belong to the second category of iconic models. The Feynman phasor model, which can be justified through didactic reasoning from quantum electrodynamics, is an example of an abstract-mathematical model. These models are structures between the quantities in physics, which cannot be observed or only with great difficulty. They

## 1. Einleitung

Das Licht in seiner komplexen physikalischen Natur ist den Menschen seit jeher ein großes Rätsel und es bereitet uns Schwierigkeiten, es zu verstehen. So sagte Einstein einmal: „Den Rest meines Lebens möchte ich damit zubringen, darüber nachzudenken, was Licht ist."

Im Laufe der Zeit wurden daher viele unterschiedliche Modelle zum Licht entwickelt, die uns helfen sollen, Licht und sein Verhalten erklären zu können: das Strahlenmodell, das Wellenmodell nach Huygens, das Teilchenmodell nach Newton, das Photonenmodell und das Zeigermodell nach Feynman. Die Modelle existieren nebeneinander und sind alle „gleich richtig". Denn ein Modell kann nicht richtig oder falsch sein, sondern nur zweckmäßig oder unzweckmäßig, da es sich um Veranschaulichungen und Vorstellungen handelt, die außerhalb der Wahrnehmung des Menschen liegen. So kann man zu einem Modell nur feststellen, ob mit ihm ein bestimmtes Phänomen gut, das heißt für den Lernenden verständlich, erklärt werden kann. So können unterschiedliche Menschen sich ein Phänomen mit unterschiedlichen Modellen erklären, sofern dies in sich schlüssig und logisch ist.

In der Schule werden zurzeit vor allem das Strahlenmodell im Anfangsunterricht gelehrt, das Wellenmodell in der Sekundarstufe I und das Photonenmodell in der Sekundarstufe II. Das Lichtteilchenmodell nach Newton und das Zeigermodell nach Feynman finden kaum Beachtung. Zum Zeigermodell gibt es bereits Untersuchungen und Unterrichtsvorschläge zur Vermittlung und Einbindung in den Unterricht der Sekundarstufe II [1]. Das Teilchenmodell wird häufig nur bei der Behandlung des Photonenmodells nebenbei eingeführt, wenn vermittelt werden soll, dass dieses nicht ausreicht, um sowohl Interferenzen als auch den Photoeffekt zu erklären.

Die fünf Modelle des Lichts lassen sich unterschiedlichen Klassen von Modellen zuordnen (siehe Abb. 1). Man unterscheidet hinsichtlich der verschiedenen Modelltypen zwischen gegenständlichen und abstrakten Modellen [2]. Erstere werden auch als konkrete Modelle oder Gebildemodelle bezeichnet und schließen zum Beispiel den Globus und Modelleisenbahnen mit ein. Die abstrakten Modelle unterteilt man in ikonische (bildhafte) Modelle und in symbolische oder abstrakt-mathematische Modelle [3]. Die ikonischen Modelle sind Erzeugnisse des menschlichen Geistes, Vorstellungen, die sich der Mensch macht von etwas Realem aber sonst unanschaulichen. Je nach Grad des Realitätsbezuges unterscheidet man die ikonischen Modelle in zwei Arten: Solche, die durch Idealisierung der Realität entstehen und solche, die nur in der Vorstellung existieren – mit einem sehr geringen Bezug zur Realität. Das Strahlenmodell des Lichts ist eine Idealisierung, das evade a viewable interpretation and are only understood as symbols of reality.

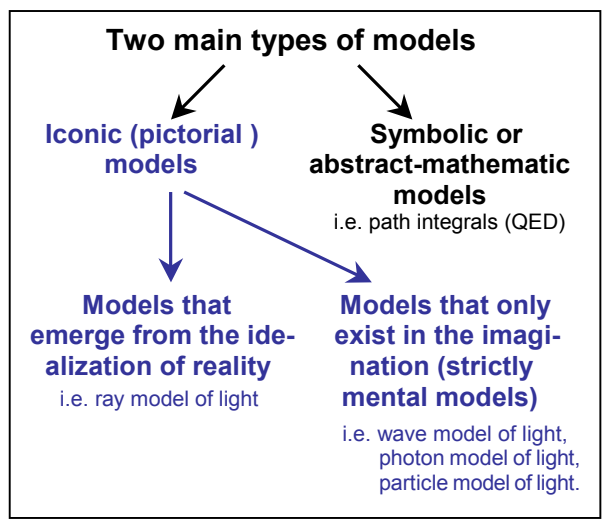

*Figure 1:* *Classification of light models into categories.*

Numerous physical phenomena can be explained with a single model. When this model is repeatedly used and anchored, as is the case with the ray model in optics taught at level I, then problems will occur if phenomena are perceived, which are no longer sensibly explainable using this model. The transition to interference appearances demonstrates such a didactic interface. In lessons concerning the analysis of experiments on interference optics, the wave model is usually referred to. Since the wave model is assigned to a different class of models than the ray model, the change in model additionally demonstrates a change in the class of model. This provides special difficulties for the students.

At the University of Potsdam three research projects have been and are being completed, which concern the process of learning with and from models. Firstly, this was the topic of a state final examination paper by Antje Leisner, in which perceptions of models of light at the upper level physics classes at the *Gymnasium* were studied using optic refraction and dispersion. Secondly, Martin Erik Horn is currently working on his dissertation concerning the development and evaluation of a curricular concept for holography, analyzing learning processes in interference optics. And finally, Antje Leisner's planned dissertation topic has as its focal point the learning process with and from models in view of the long-term development of meta-conceptual competencies in physics and chemistry lessons at level I in the *Gymnasium*. Results from the first two papers will subsequently be presented in this article.

Teilchenmodell und das Wellenmodell des Lichts gehören zur zweiten Art der ikonischen Modelle. Der Feynman'sche Zeigerformalismus, der durch didaktische Reduktion aus der Quantenelektrodynamik begründet werden kann, ist ein Beispiel für ein abstrakt-mathematisches Modell. Diese Modelle sind Realitätsgefüge zwischen physikalischen Größen, die gar nicht oder nur zum Teil unmittelbar beobachtet werden können. Sie entziehen sich der anschaulichen Deutung und sind nur als Symbole der Wirklichkeit aufzufassen.

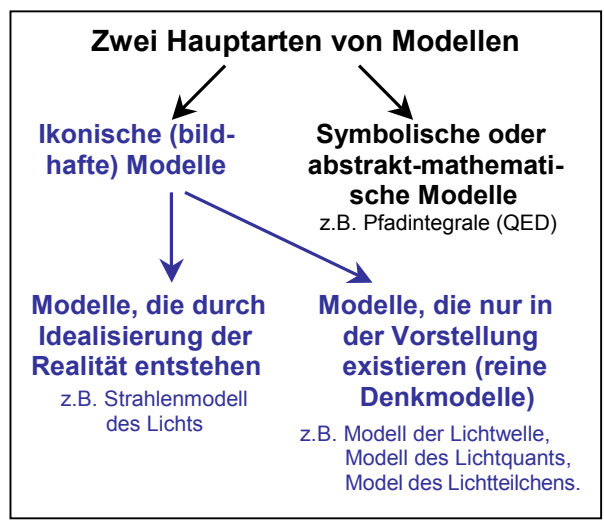

*Abbildung 1:* *Zuordnung der Lichtmodelle zu Modellklassen.*

Zahlreiche physikalische Phänomene können mit einem einzigen Modell erklärt werden. Ist dieses eingeübt und verankert, wie beispielsweise in der Optik das Strahlenmodell in der Sekundarstufe I, so treten dann Probleme auf, wenn Phänomene betrachtet werden, die mit diesem Modell nicht mehr sinnvoll zu klären sind. Der Übergang zu Interferenzerscheinungen stellt solch eine didaktische Schnittstelle dar. Im Unterricht wird zur Analyse von Experimente zur Interferenzoptik in der Regel das Wellenmodell herangezogen. Da das Wellenmodell einer anderen Modellklasse als das Strahlenmodell zugeordnet ist, stellt der auftretende Modellwechsel zusätzlich einen Modellklassenwechsel dar. Dies bereitet den Lernenden besonders Schwierigkeiten.

An der Universität Potsdam wurden und werden drei Forschungsarbeiten angefertigt, die sich mit dem Lernen mit und von Modellen beschäftigen. Das war zum Ersten die Staatsexamensarbeit von Antje Leisner zu den Modellvorstellungen zum Licht im Physikunterricht der gymnasialen Oberstufe am Beispiel der optischen Refraktion und Dispersion, zum Zweiten ist das die Dissertation von Martin Erik Horn zur Entwicklung und Evaluation eines Unterrichtskonzepts zur Holographie mit Untersuchungen von Lernprozessen zur Interferenzoptik. Und schließlich hat auch das Dissertationsvorhaben von Antje Leisner zur langfristigen Entwicklung metakonzeptueller

## 2. Use of models and processes of model construction

### 2.1 Optical dispersion and refraction

*- Goals, design and construction of case study*

There had been intense physics education research activities concerning the usage of each model of light separately. The state examination paper now explored how all five models of light are used in regard to one optic phenomenon. Based on previous studies it was assumed that students, who had just completed school had almost no meta-conceptual competencies. In brief, this means that they cannot proficiently apply the various models.

A series of lessons was developed according to the objectives of this study. This means that the curriculum was not developed for teaching at school, but was rather a didactically logical sequence of class-like sessions, conducted with six first-semester physics students. Due to the low number of participants, a quantitative evaluation was not considered.

The following questions were at the center of the case study and were used to develop the "class sessions":

1. Which concepts of light already exist?
2. How do students work with various models?
3. Where do problems occur as models are switched?
4. Do meta-conceptual phases help in the learning process?
5. What role can a computer play in the construction of models?

A knowledge test and a concept map were introduced as measuring instruments. Additionally, the lessons were videotaped, so that the students' statements could be placed in context with the tests.

The study was comprised of eight 45-minute sessions, carried out two sessions at a time. During the first two hours the Feynman phasor formalism was presented to the students. A review of the common explanation of optical dispersion and refraction followed. There emphasis was laid on the question what models are. Following this, optical refraction and dispersion were explained with the various models of light. During each lesson, the model and its characteristics were reviewed first. After that refraction was explained using the particular model in the style of a Socratic discussion. Each time an exercise followed, which the students carried out independently. For the first three models, the exercise consisted of work on the computer with the following programs: *phenOpt (Optics Phenomena)* and *Albert – The Law of Refraction*. Specifically for the phasor formalism model, work included recording resulting phasors on an accompanying worksheet. The following table displays the topic of each session and respective goals.

Kompetenzen im Physik- und Chemieunterricht der Sekundarstufe I seinen Schwerpunkt auf dem Lernen mit und von Modellen. Ergebnisse aus den ersten beiden Arbeiten werden im Folgenden vorgestellt.

## 2. Modellgebrauch und Modellbildungsprozesse

### 2.1 Optische Dispersion und Refraktion

*- Ziele, Design und Aufbau der Fallstudie*

Es gibt in der Didaktik der Physik zahlreiche Untersuchungen zu den einzelnen Modellen des Lichts. In der Staatsexamensarbeit sollten der Umgang mit allen fünf Modellen des Lichts an einem optischen Phänomen untersucht werden. Aufgrund vorhergehender Studien bestand die Annahme, dass Studierende, die gerade die Schule verlassen haben, kaum über metakonzeptuelle Kompetenzen verfügen. Das heißt, ganz kurz gesagt, dass sie nicht mit den unterschiedlichen Modellen problemgerecht umgehen können.

Für die Untersuchung wurde eine Stundenfolge entworfen, die den Zielen der Studie angepasst wurde. Das heißt, es entstand kein Curriculum für die Schule, sondern eine didaktisch sinnvolle Aneinanderreihung von unterrichtsartigen Stunden, die mit sechs Studierenden des ersten Semesters der Physik durchgeführt wurden. Da die Probandenzahl so gering war, wurde eine quantitative Auswertung von vornherein ausgeschlossen.

Die folgenden Fragen standen im Mittelpunkt der Fallstudie anhand denen die „Unterrichtsstunden" entworfen wurden:

1. Welche Konzepte vom Licht sind vorhanden?
2. Wie arbeiten die Lernenden mit unterschiedlichen Modellen?
3. Wo liegen Probleme beim Modellwechsel?
4. Helfen beim Lernen metakonzeptuelle Unterrichtsphasen?
5. Welche Rolle kann bei der Modellbildung der Computer spielen?

Als Messinstrumente wurden ein Wissenstest und ein zu bearbeitendes Concept Map eingesetzt, als auch der Unterricht vollständig videografiert, so dass auch die Studierendenaussagen mit den Tests in Zusammenhang gebracht werden konnten.

Die Studie umfasste acht Stunden á 45 Minuten und wurde jeweils in Doppelstunden unterrichtet. In den ersten zwei Stunden wird den Studentinnen und Studenten der Feynman'sche Zeigerformalismus vorgestellt. Es folgte eine Wiederholung zur allgemeinen Erklärung der optischen Dispersion und Refraktion. Zudem wurde auf Modelle in der Physik und deren Merkmale eingegangen. Im Anschluss wurde die optische Refraktion und Dispersion mit den verschiedenen Modellen des Lichts erklärt. Dabei wurden in jeder Stunde zuerst das Modell und seine Eigenschaften wiederholt. Im Anschluss wurde die Brechung mithilfe des jeweiligen Modells im sokratischen Gespräch erklärt. Dann folgte jeweils eine Übung, die die Studierenden selbständig durchführten. Sie bestand für die ersten drei Modelle in einer Arbeit am Computer mit den Programmen *phenOpt* bzw. *Albert – Brechungsgesetz* und für das Zeigermodell im Bestimmen des resultierenden Zeigers auf dem Arbeitsblatt. Die folgende Tabelle zeigt die Themen der einzelnen Stunden und das jeweilige Teilziel.

| Lesson | Theme | Partial Goal |
|---|---|---|
| A1 | Pretest, Introduction to Feynman's phasor formalism | Determine previous knowledge, set foundation for lessons |
| A2 | Phasor formalism continued | Use phasor model with reflection |
| 1. | Introduction and motivation<br>Orientation on goals<br>Models and their subdivisions<br>Repetition of refraction | Experiment by the instructor<br>Models (Definition, uses, …)<br>Review of optical refraction and dispersion |
| 2. | Geometric optics – ray model | Review of the model's characteristics<br>Derivation of the law of refraction according to Fermat, quantitative<br>Exercise using *Optics Phenomena*<br>Summary: advantages and disadvantages of the model |
| 3. | Wave optics – wave model | Review of the model's characteristics<br>Derivation of the law of refraction according to Huygens' principle<br>Explanation of dispersion<br>Exercise using *Optics Phenomena*<br>Summary: advantages and disadvantages of the model |
| 4. | Light-particle model according to Newton | Historical development<br>Review or introduction of the model<br>Explanation of dispersion and refraction<br>Exercise using *Albert*<br>Summary: advantages and disadvantages of the model |
| 5. | Feynman's phasor formalism | (Review of the model)<br>Explanation of refraction and dispersion (worksheet)<br>Summary: advantages and disadvantages of the model |
| 6. | Model comparison<br><br>Post-test | Reflection on the application of models<br>Discussion of models<br>Verification of an increase in knowledge |

*Figure 2:* *Structure of lessons.*

| Stunde | Thema | Teilziel |
|---|---|---|
| A1 | Vortest, Einführung Feynman'scher Zeigerformalismus | Vorwissen erfassen, Grundlage für Unterricht legen |
| A2 | Zeigerformalismus weiterführen | Zeigermodell am Beispiel der Reflexion anwenden |
| 1. | Einführung und Motivation<br>Zielorientierung<br>Modelle und deren Unterteilung<br>Wiederholung Brechung | Lehrerexperiment<br>Modelle (Definition, Anwendungen, ...)<br>Wiederholung der optischen Refraktion und Dispersion |
| 2. | Geometrische Optik – Strahlenmodell | Wiederholung der Merkmale des Modells<br>Herleitung Brechungsgesetz nach Fermat, quantitativ<br>Übung mit *phenOpt*<br>Zusammenfassung: Vor- und Nachteile des Modells |
| 3. | Wellenoptik – Wellenmodell | Wiederholung der Merkmale des Modells<br>Herleitung Brechungsgesetz nach dem Huygens' schen Prinzip<br>Erklären der Dispersion<br>Übung mit *phenOpt*<br>Zusammenfassung: Vor- und Nachteile des Modells |
| 4. | Lichtteilchenmodell nach Newton | Historische Entwicklung<br>Wiederholung bzw. Einführung des Modells<br>Erklärung der Dispersion und Refraktion<br>Übung mit *Albert*<br>Zusammenfassung: Vor- und Nachteile des Modells |
| 5. | Feynman'scher Zeigerformalismus | (Wiederholung des Modells)<br>Erklären der Refraktion und Dispersion (Arbeitsblatt)<br>Zusammenfassung: Vor- und Nachteile des Modells |
| 6. | Vergleich der Modelle<br><br>Nachtest | zum Umgang mit Modellen reflektieren<br>Diskussion der Modelle<br>Überprüfen des Lernzuwachses |

***Abbildung 2:*** *Struktur des Unterrichts.*

*- Follow-Through and Testing*

The independent activities of the students as well as the Socratic discussions for the clarification of the models and their application with optical refraction and dispersion were placed at the center of the study. An explanation of the refraction using the particle model of light is presented as an example (see figure 3).

*- Durchführung und Tests*

Bei der Durchführung der Studie standen die selbständigen Tätigkeiten der Studierenden und die sokratischen Gespräche zur Erarbeitung der Modelle und deren Anwendung am Beispiel der optischen Refraktion und Dispersion im Vordergrund. Als Beispiel wird hier kurz die Erklärung der Refraktion im Lichtteilchenmodell vorgestellt (siehe Abbildung 3).

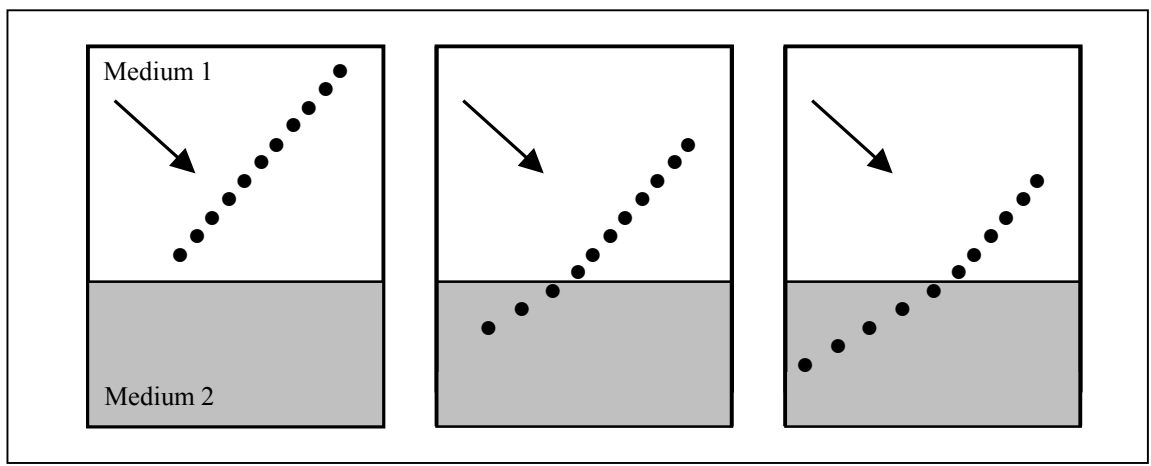

***Figure 3:*** *Optical refraction in the particle model of light.*
***Abbildung 3:*** *Optische Refraktion im Lichtteilchenmodell.*

The particle chain shown in figure 3 moves through medium 1 at speed $c_1$. The first particle in the chain meets with medium 2 and because this substance is optically more dense, the speed $c_2$ in medium 2 will be lower than $c_1$ in medium 1. The remaining particles in medium 1 will continue to move with the original speed, until they enter into medium 2 as well and are "slowed down." This leads to a bend in the direction of movement – to the breaking of the particle chain.

During this explanation a student commented (after refraction had previously been explained using the ray model and wave model): "Now I understand it, the particles in the air cover a greater distance [in the same amount of time] than the particles in the water." Information regarding the application and understanding of models could also be obtained during the use of other models.

This information was then compared to the results of the test. In a knowledge test, students did not name the particle model as a model of light, but in the concept map it does become evident that they also perceive light as particles. In the knowledge test as well as in an additional post-test, further problems and incorrect perceptions are noticeable, as illustrated in question 6 (figure 4). The students know that refraction of light plays a role, yet they "confuse" the direction of light, which provides an individual with vision. One student's sketch (figure 4) and his description clearly show that he in-

Die in Abbildung 3 dargestellte Teilchenkette bewegt sich im Medium 1 mit der Geschwindigkeit $c_1$, das erste Teilchen der Kette trifft auf das Medium 2 und da dies optisch dichter ist, ist die Geschwindigkeit $c_2$ im Medium 2 kleiner als $c_1$ im Medium 1. Die übrigen Teilchen im Medium 1 bewegen sich mit der ursprünglichen Geschwindigkeit weiter, bis sie ebenfalls in das Medium 2 eintreten und „abgebremst" werden. Dies führt zum Abknicken der Bewegungsrichtung – zur Brechung der Teilchenkette.

Bei dieser Erklärung äußerte ein Student, nachdem die Refraktion bereits im Strahlenmodell und Wellenmodell erklärt worden war: *„Jetzt habe ich das verstanden – die Teilchen in der Luft legen eine größere Strecke* [in der gleichen Zeit] *zurück als die Teilchen im Wasser."* Auch bei der Behandlung der anderen Modelle konnten aus den Probandenaussagen Informationen zum Modellgebrauch und Modellverständnis gewonnen werden.

Diese wurden dann mit den Ergebnissen des Tests verglichen. Im Wissenstest benennen die Studierenden im Vortest das Teilchenmodell nicht als Modell des Lichts, aber im Concept Map wird beispielsweise deutlich, dass sie sich Licht auch als Teilchen vorstellen. Zudem sind im Wissenstest wie auch im Nachtest weiterhin Probleme und Fehlvorstellungen erkennbar, wie die Aufgabe 6 (siehe Abb. 4) zeigt. Die Studierenden wissen, dass die Lichtbrechung eine Rolle spielt, „verwechseln" aber die Richtung des Lichtes, das für das jeweilige directly assumes the dispersion of light from the eye to an object, yet disregards the diffusion on the object.

Sehen der Personen sorgt. Die Skizze eines Studenten (Abb. 4) und seine Erklärung zeigen deutlich, dass er indirekt von einer Lichtausbreitung vom Auge zum Objekt ausgeht und die Streuung am Objekt missachtet.

**Question 6:** *Explain the following with the help of a sketch. Person A stands several meters away from the shore. Can he see person B underwater? Can person B see person A on the shore? Explain your reasoning with a sketch!*

**Aufgabe 6:** *Der Mensch A steht mehrere Meter vom Ufer entfernt. Kann er den tauchenden Menschen B im Wasser sehen? Kann der Mensch B den Menschen A am Ufer sehen? Erklären Sie die Vorgänge mit Hilfe einer Skizze!*

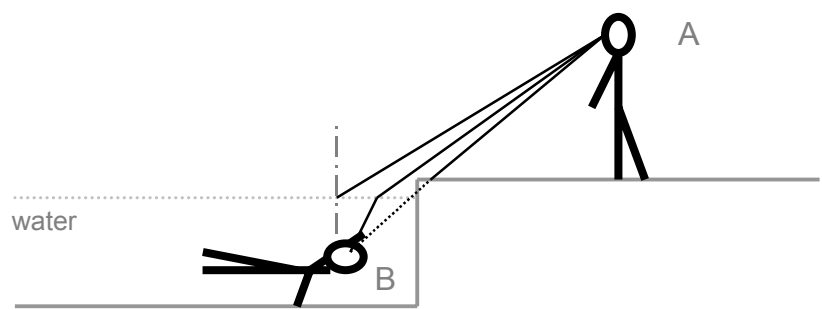

*"A can see B, because the light on the water's surface is broken. If the light passes into an optically denser medium, it breaks at an angle. B can also see A, since the light again breaks away upon exiting."*

*"A kann B sehen, da das Licht an der Wasseroberfläche gebrochen wird. Wenn das Licht in ein optisch dichteres Medium übergeht, wird es zum Lot hin gebrochen. B kann A auch sehen, da das Licht beim Austritt vom Lot weg gebrochen wird."*

*Figure 4:* Student's explanation and illustration of q*uestion 6.*
*Abbildung 4: Aufgabe 6 mit eingezeichnetem Strahlengang und Erklärung des Studenten.*

Overall students demonstrated a distinct increase in knowledge regarding the use of models after participating in the lessons. Additionally, results to the questions of the case study were determined as follows.

### - Results

1. Which concepts of light are available?

Students bring concrete, yet sometimes incorrect, ideas about light from school. These erroneous perceptions of models are often very strong. This is how the students explain differing phenomena of light in the ray model and in the wave model and identify the wave-particle dualism. However, they are not able to explain or employ the dualism. They are not familiar with the particle model of light according to Newton, although perceptions of this type of model were noticeable in the concept maps.

2. How do students work with various models?

The test students hardly use the models of light consciously. They use the terms "light", "light ray",

In der Gesamtheit zeigen die Studierenden nach dem Unterricht einen deutlichen Wissenszuwachs im Umgang mit Modellen und es konnten folgende Ergebnisse auf die Untersuchungsfragen gefunden werden.

### - Ergebnisse

1. Welche Konzepte vom Licht sind vorhanden?

Die Studierenden bringen aus der Schule konkrete, doch zum Teil falsche, Vorstellungen zum Licht mit. Diese sind zu den einzelnen Modellen sehr gefestigt. So erklären sie unterschiedliche Phänomene des Lichts im Strahlenmodell und im Wellenmodell und benennen den Welle-Teilchen-Dualismus. Anwenden können sie den Dualismus zur Erklärung nicht. Das Lichtteilchenmodell nach Newton ist ihnen nicht bekannt, obwohl in den Concept Maps Vorstellungen dieser Art deutlich werden.

2. Wie arbeiten die Lernenden mit unterschiedlichen Modellen?

Die Probanden gehen kaum bewusst mit den Modellen des Lichts um. Sie benutzen die Begriffe

"particle" and "wave" almost synonymously. Hence, confusion between reality and model arises.

3. Where do problems occur as models are switched?

Students have positive experiences in using models; this unfortunately lies in the high clarity of the models. The students for this reason have on the one hand an open attitude towards the learning of the phasor formalism, yet, on the other they struggle with the high degree of abstraction. Consequently, when learning a new model or when switching to a different model, students attempt to incorporate the new with the old.

4. Do meta-conceptual phases help in the learning process?

As a consequence of meta-conceptual phases in teaching, in which the formation, application and relationship to other models of light were covered, light in its complex physical nature was becoming more comprehensible to the tested persons. As a consequence, what is model and what reality is not confused as often, since the model and its characteristics are applied purposely. This increases the understanding for the work with models and with the respective phenomenon.

5. What role can a computer play in the construction of models?

Computers can aid in the visualization of models. The dynamic components of models in particular (the propagation of waves, superposition and interference of waves, velocity of light in different media) allow themselves to be presented very well through computer simulation programs. Exercises in which the students use the particular model itself, support the learning process, since only activities of the students who should handle the models can improve the ability to deal with models in a problem-solving manner.

**2.2 Interference optics: holography**

The second focal point that the University of Potsdam is working on in the field of model construction research is interference optics. Students who participated in a course on holography were questioned about the application and development of models within the framework of a dissertation. The series of courses was taught at a total of five Brandenburg secondary schools (*Gymnasium*) at the 13$^{th}$ grade level. The structure of this course and first results were already presented in [5] and [6]. 54 students took part in this research project.

„Licht", „Lichtstrahl", „Teilchen" und „Welle" nahezu synonym. Daher entstehen Vermischungen von Wirklichkeit und Modell.

3. Wo liegen Probleme beim Modellwechsel?

Die Studierenden haben positive Erfahrungen im Umgang mit Modellen gemacht: Dies liegt leider an der hohen Anschaulichkeit der Modelle. Aus diesem Grunde haben sie zum einen beim Erlernen des Zeigerformalismus eine offene Einstellung und tun sich aber zum anderen schwer mit dem hohen Maß an Abstraktion. Sie versuchen daher beim Modellwechsel bzw. dem Erlernen von neuen Modellen das Neue in das Alte einzubinden.

4. Helfen beim Lernen metakonzeptuelle Unterrichtsphasen?

Aufgrund der metakonzeptuellen Unterrichtsphasen, in denen Entstehung, Anwendung und die Zusammenhänge zu anderen Modellen des Lichts behandelt wurden, wird den Probandinnen und Probanden das Licht in seiner komplexen physikalischen Natur zunehmend besser bewusst. Zudem kommt es weniger zu Vermischungen von Wirklichkeit und Modell, da das Modell und seine Merkmale bewusst benutzt wurden. Dies erhöht das Verständnis für das Arbeiten mit Modellen und damit des jeweiligen Phänomens.

5. Welche Rolle kann bei der Modellbildung der Computer spielen?

Der Computer kann der Veranschaulichung der Modelle dienen. Gerade die dynamische Komponente der Modelle (Fortpflanzen der Wellenberge, Überlagerung und Auslöschung, Änderung der Ausbreitungsgeschwindigkeit in unterschiedlichen Medien) lässt sich in Computersimulationsprogrammen sehr gut darstellen. Übungen, in denen die Lernenden das jeweilige Modell selbst anwenden, sind lernunterstützend, da nur durch das Durchführen von Tätigkeiten die Fähigkeit erlangt werden kann, mit Modellen problemgerecht umzugehen.

**2.2 Interferenzoptik: Holographie**

Der zweite an der Universität Potsdam bearbeitete Schwerpunkt im Bereich der Modellbildungsforschung ist die Interferenzoptik. Im Rahmen einer Promotionsarbeit wurden Modellgebrauch und Modellentwicklung von Schülerinnen und Schülern untersucht, die an einem Unterricht zur Holographie teilgenommen haben. Die Unterrichtsreihe wurde an insgesamt fünf Brandenburger Gymnasien in der 13. Jahrgangsstufe unterrichtet. Ihr Aufbau und erste Ergebnisse wurden bereits in [5] und [6] vorgestellt. Die Stichprobengröße betrug 54 Schülerinnen und Schüler.

The objectives of the study were:

- development, test and evaluation of an educational concept for holography, which would be applied at the upper level of secondary schools.
- assessment of the students' perception of holography and interference optics under the special consideration of model construction processes.

This paper will mainly consider the second objective.

*- Study overview*

The following instruments of evaluation were implemented: questioning of the students with pre-, post- and long-term tests as well as concept maps, videotaping of the class and conducting interviews with students. In order to examine how models are comprehended and applied by the students, the last two focal points are considered in more detail.

Since the results of the first objectives were presented earlier, we will provide only a brief overview here. The analysis of the results (first objective) demonstrated that the knowledge of the group of students concerning the basics of holography increased considerably. As it is assumed that the students have a low familiarity with the subject, this increase is highly significant and demonstrates a successful learning process.

The level of prior knowledge of the students concerning geometric optics and the wave phenomenon was more advanced. Nevertheless, highly significant effects of comprehension were noticeable, even if the increase in knowledge was not as high as in the case of knowledge concerning holography. This was to be expected, since the topic of the lecture was holography. The established small successes in learning in the other areas can be considered as positive side effects and also illustrate the networked and already existing structure of the teaching series.

*- Evaluation of student interviews*

A typical conclusion in the evaluation of interviews conducted with individual students at the end of the course highlights the lack of differentiation between the models of light. The spectrum of the discovered perspectives reaches from the partial overlapping of models to the integration of one model into another and to the complete hybridization of all models known by the student, as exemplified by figure 5.

This sketch was drawn by a student in a participating physics course, in order to illustrate how he imagines the nature of light. It is obvious that he incorporates the three models covered in the course. The ray model serves the student as a foundation and cognitive anchor. In order to expand it consis-

Die Ziele der Untersuchung waren:

- Entwicklung, Erprobung und Evaluation eines Unterrichtskonzepts zur Holographie für den Einsatz in der gymnasialen Oberstufe.
- Untersuchung von Schülervorstellungen zur Holographie und zur Interferenzoptik unter besonderer Berücksichtigung von Modellbildungsprozessen.

In diesem Beitrag soll schwerpunktmäßig über die zweite Zielstellung berichtet werden.

*- Überblick über die Studie*

Es wurden die folgenden Evaluationsinstrumente eingesetzt: Anfertigung von Vor-, Nach- und Langzeittests sowie von Concept Maps durch die Lernenden, Videographie des Unterrichts und Durchführung von Schülerinterviews. Zur Untersuchung des Modellverständnisses und des Modellgebrauchs der Schülerinnen und Schüler wurden schwerpunktmäßig die beiden letztgenannten ausgewertet.

Da über die Ergebnisse der ersten Zielstellung bereits berichtet wurde, hier nur ein kurzer Überblick: Die Wirkungsuntersuchung (erste Zielstellung) ergab, dass das Wissen der Lerngruppen über die Grundlagen der Holographie deutlich angestiegen ist. Ausgehend von geringem Vorwissen ist dieser Anstieg höchst signifikant und belegt gute Lernerfolge.

Das Vorwissen der Schülerinnen und Schüler in den Bereichen der geometrischen Optik und der Wellenphänomene lag auf einem höheren Niveau. Es konnten auch hier höchst signifikante, wenn auch nicht so starke, Lerneffekte gezeigt werden. Dies war zu erwarten, da der Unterricht sich auf das Thema der Holographie bezog. Die nachgewiesenen geringeren Lernerfolge in den anderen Bereichen können als positive Nebeneffekte angesehen werden und zeigen den vernetzten, sich auf bereits vorhandene Strukturen stützenden Aufbau der Unterrichtsreihe.

*- Auswertung der Schülerinterviews*

Ein typischer Befund bei Auswertung der im Anschluss an die Unterrichtseinheit mit einzelnen Schülerinnen und Schülern aller Lerngruppen geführten Interviews stellt die fehlende Trennschärfe zwischen den Modellen zum Licht dar. Das Spektrum der aufgefundenen Vorstellungen reicht dabei von teilweisen Überlappungen der Modelle über die Integration eines Modells in ein anderes bis hin zur vollständigen Hybridisierung aller dem Schüler bekannten Modelle, wie dies beispielsweise Abbildung 5 zeigt.

Diese Skizze wurde von einem Schüler eines beteiligten Physikleistungskurses angefertigt, um zu verdeutlichen, wie er sich die Natur des Lichtes vorstellt. Deutlich zu erkennen ist die Einbindung tently in his view, he postulates an enlargement of the ray of light. By characterizing the light ray as a medium for the motion of waves, the student generates a unification of models in this newly created space of perception, which was naturally not part of the lesson. The wave itself is described as a stimulated particle field in a naïve particle perspective, which is similar to a chain.

In the remainder of the interviews, the psychological advantages of these hybrid experiences become evident. Provided that the surface structure is con-der drei im Unterricht behandelten Modelle. Das Strahlenmodell dient dem Schüler als Grundlage und kognitiver Anker. Um es in seiner Sicht konsistent zu erweitern, postuliert er eine Vergrößerung des Lichtstrahls. In diesem neu geschaffenem Vorstellungsraum, der im Unterricht naturgemäß nicht besetzt wurde, nimmt der Schüler seine Vereinheitlichung vor, indem er den Lichtstrahl als Träger einer Wellenbewegung charakterisiert. Die Welle selbst wird in naiver Teilchenvorstellung als bewegtes Teilchenfeld, einer Kette nicht unähnlich, beschrieben.

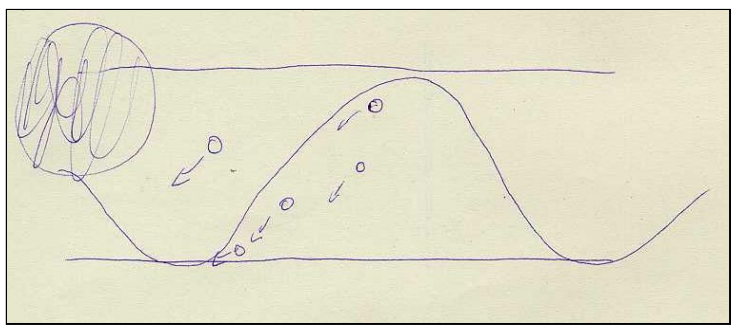

**Student:** Actually, one doesn't specify a model, but rather confuse the model perspectives. So there basically existed a basic theory, lets say, with the ray optics and with the ray model. Then one discovered that this… actually this little ray is called a ray and principally this straight line, if one zooms really up close, could be something like a wave (…) And if I get a bit closer then I could imagine for example, that somewhere in this wave, the wave consists of some sort of particles.

**Schüler:** Eigentlich ist es so, dass man nicht ein Modell verfeinert, sondern dass man die ganzen Modellvorstellungen eher vermischt. Dass man im Grunde so eine Grundtheorie, sag ich mal, hatte, mit der Strahlenoptik, mit dem Strahlenmodell. Dass man dann darauf gekommen ist, dass ja dieser kleine Strahl, Strahl heißt ja im Grunde genommen eine gerade Linie, wenn man die ganz doll ranzoomt, so eine Art Welle sein könnte. (...) Und wenn ich noch weiter rangehe, dann könnte ich mir zum Beispiel vorstellen, dass irgendwo in dieser Welle da, die Welle aus irgendwelchen Teilchen besteht.

*Figure 5:* Drawing of a student and interview excerpt I.
*Abbildung 5:* Schülerzeichnung und Interviewausschnitt I.

gruent to the phenomenon to be explained and to the selected model, the student can alter the perspective of the model at will, move back and forth between a number of perspectives or can selectively combine in order to create patterns of explanation that fit at the time of implementation. If there are contradictions within a model then it is consequently transferred to another model without questioning whether the selected patterns of explanation are physically possible.

Figure 6 shows such a pattern of explanation of the same student, who was trying to describe the behavior of light using double slits in this sketch. He draws rays of light originating from the light source, illustrated here as the sun, which are positioned in the sense of a far field description. Just in front of the double slits he changes the proportions and simultaneously the conception of the model. Now the student argues based on the photon model, which he in turn presents as a naïve conception of a particle. Since he cannot explain the maxima of the

Im weiteren Verlauf des Interviews werden die psychologischen Vorteile dieser Hybridanschauungen deutlich: Der Schüler kann die Modellvorstellungen beliebig wechseln, zwischen mehreren Modellvorstellungen pendeln oder selektiv vermischend gerade passende Erklärungsmuster erstellen, sofern ihm die Oberflächenstruktur von dem zu erklärenden Phänomen und dem ausgewählten Modell kongruent erscheint. Zeigen sich Widersprüche in einer Modellvorstellung, wird konsequent in eine andere Modellvorstellung übergegangen, ohne zu hinterfragen, ob die gewählten Erklärungsmuster physikalisch sinnvoll sind.

Abbildung 6 zeigt ein solches Erklärungsmuster des gleichen Schülers, der mit Hilfe dieser Skizze das Verhalten von Licht am Doppelspalt zu erläutern sucht. Von der Lichtquelle, hier dargestellt als Sonne, skizziert er Lichtstrahlen, die im Sinne einer Fernfeldbeschreibung angesetzt werden. Kurz vor dem Doppelspalt wechselt er die Größenverhältnisse und gleichzeitig die Modellvorstellung. Jetzt ar-

> **Question:** (...) Why will the light actually hit these spots?
>
> **Student:** Well, basically if one assumes that maximums really exist here, then there must be a superposition of photons – somewhere, or from light that oscillates and produces maximums and in between at other points there must then basically be a deterioration of light. This means that somehow the photons, which for example exit here, must also arrive here behind the slits. So, they have to somehow scatter here a little bit.
>
> 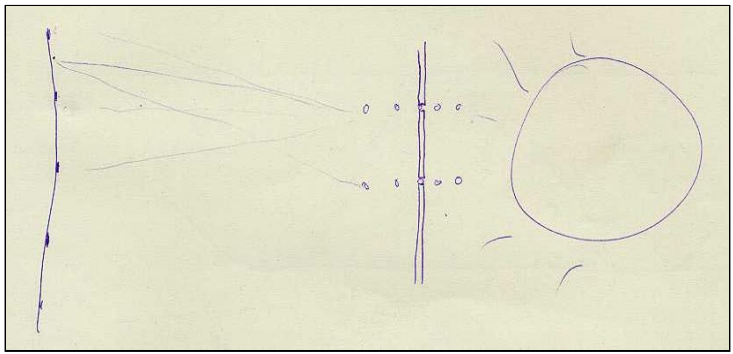
>
> **Frage:** (...) Weshalb wird tatsächlich an diesen Stellen Licht auftreffen?
>
> **Schüler:** Ja, im Grunde wenn man davon ausgeht, dass das hier wirklich Maxima sind, muss das ja eine Überlagerung von Photonen sein - irgendwo oder von Licht, das sich aufschaukelt und Maxima erzeugt und zwischendurch an den anderen Punkten muss ja dann im Grunde eine Abschwächung sein. Das heißt, irgendwie müssen ja hier hinter dem Spalt sich die Photonen, die meinetwegen hier rauskommen, müssen ja auch hier ankommen. Also die müssen irgendwie hier so ein bisschen streuen.

*Figure 6: Drawing of a student and interview excerpt II.*
*Abbildung 6: Schülerzeichnung und Interviewausschnitt II.*

interference pattern, he automatically jumps into a modified illustration of a ray. The light changes its character in the middle of the space. Similar to the demonstration of light sources in physics class, which are usually illustrated on the chalkboard as a source emanating rays in numerous directions, the student draws rays that emanate from particles. He simultaneously chooses an illustration that is selective in direction, which prevents the light rays from hitting the screen evenly. In the interview he meanwhile speaks of "superposition" and "oscillation", where he exemplifies that he did not deviate from the wave model during his entire explanation.

Thus the student does not only link the conceptions of the model to an all-encompassing hybrid-model, but also works with several models in a cognitive and synchronous way. The missing distinction between the model conceptions has to thus be considered one of the most important causes for a failure of attempts of explanation of the test student in optics.

In its most extreme development, the fusion of model conceptions goes as far as loosing conceptual limitations. For students of this type, a light ray, a light wave, light particles or light quanta are one and the same and are applied in a similar sense.

gumentiert der Schüler im Photonenmodell, das er wiederum naiv als Teilchenvorstellung präsentiert. Da er die Maxima des Interferenzmusters nicht erklären kann, springt er unvermittelt in eine modifizierte Strahlendarstellung: Mitten im Raum wechselt das Licht seinen Charakter. Ähnlich wie bei der Darstellung von Lichtquellen im Physikunterricht, die sich an der Tafel oft als Quelle von in zahlreiche Richtungen ausgehenden Strahlen dargestellt finden, zeichnet der Schüler von Teilchen ausgehende Strahlen. Gleichzeitig wählt er eine richtungsselektive Darstellung, die verhindert, dass die Lichtstrahlen gleichmäßig auf den Schirm treffen. Im Interview spricht er währenddessen von „Überlagerung" und „aufschaukeln", wodurch er zeigt, dass das Wellenmodell während der gesamten Argumentation nicht verlassen wurde.

Der Schüler verbindet die Modellvorstellungen somit nicht nur zu einem allumfassenden Hybridmodell, sondern arbeitet kognitiv synchron mit mehreren Modellen. Die fehlende Trennschärfe zwischen den Modellvorstellungen muss deshalb als einer der wichtigsten Gründe für ein Scheitern von physikalischen Erklärungsversuchen der Probanden angesehen werden.

In ihrer extremsten Ausprägung geht die Vermengung von Modellvorstellungen so weit, dass die begriffliche Eingrenzung völlig verloren geht. Für solche Schülerinnen und Schüler sind Lichtstrahl, Lichtwelle, Lichtteilchen oder Lichtquant ein und dasselbe und werden sinngleich verwendet.

It becomes evident that the model conceptions of numerous students are associated with a strict hierarchy. As the following interview excerpt (figure 7) shows, the student perceives the model of rays as contained in the wave model and the wave and ray model contained in the quantum model.

In this way, he does not consider new structures of models as independent constructs, but as mere expansions of previously known patterns of explanation. From the view of the student the previously covered models represent models with a lower capacity for explanation in the hierarchy of models. For this reason the model introduced last in the course, i.e. the quantum model, thus practically receives the most omnipotent position.

> **Student:** The **ray model** is really only a model, where one can describe only particular characteristics. The **wave model** would explain interference, the diffraction, … hm, what else? When the waves superpose, that cancellation and such are created. And the **quantum model** can explain that the light does interfere, but not evenly. The quanta always emerge where the quanta move along a wave, this would explain the characteristics of waves; and along the ray, with that one can explain diffraction, eh… refraction and such things.
> **Question:** ... I didn't quite understand that.
> **Student:** The **wave model** also explains the characteristics of the rays and the **quantum model** also the characteristics of the **wave model and the ray model**.
> **Question:** This means that the quantum model encompasses the other two models?
> **Student:** Yes, it is therefore better, it is closer to reality.

*Figure 7: Interview excerpt III showing the unawareness of limits of the models by the student.*

The outstanding position of the quantum model in the eyes of the students is legitimated in that the quantum model is closer to reality. This claim is based on the assumption that reality can indeed be recognized and has in fact been recognized by the teacher or other experts. As mentioned in the introduction, this is not only a problematic point of view in the case of light.

A more detailed analysis of the interview excerpt reveals a second focal point of the formation and development of models that was observed not only in the case of this student. In the first part, the illus-tration is limited to the contents conveyed in class. Here the models are described in a correct and correctly separated manner.

After this first part, which has been derived from the knowledge at disposal, the student interprets the model, transcending the presentation in class. This division is a strong indication that (as is the case with this student) the application of models in class is limited to the mere transmission of subject matter or the interpretation of models was unclear and the student unable to differentiate between reality and the particular levels of models. Therefore typical scenes from class will be analyzed in the following section.

*- Typical scenes from class*

The concept [5] of the lectures based on holography calls for a thorough discussion of the models in numerous phases of the lectures. Despite of the uniform peripheral stipulations and the predetermined concepts, the actual structure of the lectures differed in detail in all five test groups since the participating teachers, as expected, followed their own lecturing style. Moreover, not only did the discussions in class and their phases feature different focal points, but also strongly differing hidden or openly presented epistemological positions of the teachers.

Thus the plain establishment by the teacher when deliberating the concept of "coherence" (figure 8) can cause problematic conceptions.

> **Teacher:** But each photon in itself actually is – and there of course exist different illustrations of model – a large wave train.

*Figure 8: Teacher comments on the nature of light.*

Here the nature of the particles and the nature of the waves of light was equated. Since the verb "to be" (*sein*) has a strong constituting effect in the German language, the statement that "each photon is a large wave train" has a defining character and prevents a physically logic interpretation. The teacher is aware of this and thus moderates his proclamation in two ways. Firstly, the teacher reduces the effect of his proclamation by inserting the word "actually", yet the aim of the moderation remains open for the students. Does the moderation of "each" relate to the number of the affected photons, the photons themselves or to the aspects of equating the photons and waves? The teacher tries to limit this openness to the level of the model by inserting a second supplement, that there are "naturally also different model-like illustrations." In fact the likelihood exists that students will not only become uncertain, but also that they experience the mixture and equation of models as a legiti-dung und Modellentwicklung, der nicht nur bei diesem Schüler beobachtet werden konnte: Im ersten, aufzählenden Teil beschränkt sich die Darstellung auf die im Unterricht vermittelten Inhalte. Hier werden die Modelle korrekt und korrekt getrennt beschrieben.

Nach diesem ersten, dem Verfügungswissen entnommenen Teil geht der Schüler interpretierend über die unterrichtliche Darstellung hinaus. Diese Zweiteilung ist ein starkes Indiz dafür, dass (wie bei diesem Schüler) der Umgang mit Modellen im Unterricht auf die reine inhaltliche Vermittlung beschränkt blieb oder die Interpretation von Modellen unscharf und nicht zwischen den einzelnen Modellebenen und der Realität trennend erfolgte. Deshalb sollen im folgenden Abschnitt charakteristische Unterrichtsszenen analysiert werden.

*- Charakteristische Unterrichtsszenen*

Das dem Unterricht zur Holographie zugrunde liegende Konzept [5] sieht eine vertiefte Modelldiskussion in zahlreichen Unterrichtsphasen vor. Trotz der einheitlichen Randbedingungen und des vorgegebenen Konzepts fiel die tatsächliche Unterrichtsgestaltung in den fünf Untersuchungsgruppen im Detail sehr unterschiedlich aus, da die beteiligten Lehrerinnen und Lehrer naturgemäß den ihnen eigenen Unterrichtsstil verfolgten. Darüber hinaus zeigten nicht nur Unterrichtsgespräch und Diskussionsphasen unterschiedliche Schwerpunktsetzungen, sondern auch stark differierende untergründig vorhandene bzw. offen vorgetragene erkenntnistheoretische Positionen der Unterrichtenden.

So kann bereits die einfache Feststellung eines Lehrers bei Erläuterung des Begriffes der „Kohärenz" (Abbildung 8) problematische Vorstellungen auslösen.

> **Lehrer:** Aber jedes Photon für sich ist eigentlich – und da gibt es natürlich auch modellmäßig verschiedene Darstellungen – ein großer Wellenzug.

*Abbildung 8: Lehreräußerung zur Lichtnatur*

Hier werden die Teilchennatur und die Wellennatur des Lichtes gleichgesetzt. Da das Verb „sein" im Deutschen eine stark konstituierende Wirkung aufweist, hat die Aussage „Jedes Photon ist ein großer Wellenzug" definierenden Charakter und unterbindet eine physikalische sinnvolle Interpretation. Dies weiß auch der Lehrer und schwächt deshalb seine Aussage in zweifacher Hinsicht ab: Erstens vermindert er durch Einfügung des Wortes „eigentlich" die Wirkung seiner Aussage, wobei die Zielrichtung der Abschwächung für die Schüler noch offen bleibt. Bezieht sich die Abschwächung auf „jedes", also die Anzahl der betroffenen Photo-mate physical pattern of argument. Numerous contributions by the students in class show that such patterns of argument have developed. They are not always solved in class.

In addition, the basics of discussion concerning the separation and distinction of models and differentiation between the world of models and reality are not recorded, not considered or often changed as the following excerpt of a class (figure 9) illustrates.

> **Teacher:** So what is the light of a laser? Hans?
> **Student:** Well, it is very strongly bundled so that it takes on the form of a ray.
> **Teacher:** Then it is not a ray of light?
> **Student:** Well, yes, it is a ray of light. I am not certain if it is a 100 percent light ray, but it is very strongly bundled.
> **Teacher:** This is my exact question, is a laser …
> **Student:** It is very strongly bundled light and not a ray of light.
> **Teacher:** And why couldn't it be a ray of light?
> **Another Student:** Because a ray of light is a model.
> **Teacher:** Very good. At this point…
> **(distributes worksheets)**

*Figure 9: Excerpt of class concerning model discussions.*

At the beginning the teacher asks for the characteristics of a laser and thus is on the level of real manifestations. Initially the student maintains this level by talking of the intense bundling. In the second clause he however changes over to the level of the ray model, which he deduces from the real observation of an intense bundling and thus he intermingles reality and the model. The teacher not only comprehends this change in levels, but also consequently sustains this change by referring to the ray model in his question. However, the change in levels is not comprehended by the student. Even though he responds to the leading question of the teacher by arguing with the sought after term "ray of light", he has however not yet departed from the level of real manifestations. He very much limits his statement by stating that he is not sure if it is a 100 percent ray of light and again refers to the intense bundling of a laser. With this argument he remains at the level of reality. The teacher also follows the student's retreat in that he asks a question concerning the laser. The student in turn picks up by merging the levels: It is a intensely bundled light and not a ray of light. Since he cannot provide reasons for his proclamation, the student is conse-nen, auf die Photonen an sich oder auf den Aspekt der Gleichsetzung von Photon und Welle? Diese Offenheit versucht der Lehrer durch den zweiten Einschub, dass es „natürlich auch modellmäßig verschiedene Darstellungen" gäbe, auf die Modellebene zu verengen. Tatsächlich besteht durch solche Aussagen nicht nur die Gefahr, dass die Lernenden verunsichert werden, sondern ebenso, dass sie Modellmischungen und Modellgleichsetzungen als legitime physikalische Argumentationsmuster erleben. Zahlreiche Unterrichtsbeiträge der Schülerinnen und Schüler zeigen, dass sich solche Argumentationsmuster tatsächlich ausgebildet haben. Sie werden nicht immer im Unterricht geklärt.

Darüber hinaus werden die Diskussionsgrundlagen bezüglich Modelltrennung, Modellunterscheidung und Differenzierung zwischen Modell- und Realwelt teilweise nicht erfasst, nicht beachtet oder wie der folgende Unterrichtsausschnitt (Abbildung 9) zeigt, oft gewechselt.

> **Lehrer:** Was ist also Laserlicht? Hans?
> **Schüler:** Ja, also sehr stark gebündelt, dass es ja Strahlenform annimmt.
> **Lehrer:** Also kein Lichtstrahl?
> **Schüler:** Ja, doch, ein Lichtstrahl. Ich weiß nicht, ob es ein hundertprozentiger Lichtstrahl ist, aber sehr stark gebündelt.
> **Lehrer:** Das ist jetzt genau meine Frage, ist Laserlicht ....
> **Schüler:** Das ist sehr stark gebündeltes Licht und nicht ein Lichtstrahl.
> **Lehrer:** Und warum kann das kein Lichtstrahl sein?
> **Anderer Schüler:** Weil Lichtstrahl ein Modell ist.
> **Lehrer:** Sehr schön. Gut an der Stelle...
> **(teilt Arbeitsbögen aus)**

*Abbildung 9: Unterrichtsausschnitt zur Modelldiskussion*

Der Lehrer fragt zu Beginn nach den Eigenschaften von Laserlicht und befindet sich damit auf der Ebene realer Erscheinungen. Der Schüler behält diese Ebene zuerst bei, indem er auf die starke Bündelung zu sprechen kommt. Im zweiten Halbsatz wechselt er jedoch in die Modellebene des Strahlenmodells, die von ihm aus der realen Beobachtung einer starken Bündelung abgeleitet wird und so eine Vermengung zwischen Modell und Realität zeigt. Der Lehrer vollzieht diesen Ebenenwechsel nicht nur nach, sondern führt ihn konsequent fort, indem er sich durch seine Frage vollständig auf das Strahlenmodell bezieht. Dieser erfolgte Ebenenwechsel ist dem Schüler allerdings nicht klar. Obwohl er auf die Suggestivfrage des Lehrers mit dem quently not aware of the articulated change in level made by him. Only a second student in the end denotes the ray of light as a model, at which point the teacher continues with the lecture.

The problems illustrated by this excerpt of a lecture have far reaching consequences for the utilization of models. Thus the merging of models and the development of hybrid perspectives are not only and not always explainable through unreflected applications of models. The intermingling of these models can be directly explained by many students by way of observing an equation of reality and model or by the advance of the model toward reality (see figure 10).

> **Student:** ... then a model is created, well, a perception, this is the way it could be. And this is a model in my opinion. Well, and this cannot be equated with reality, that it really is like this. And, well, a model is not a model until it can definitely be proven.

*Figure 10: Comment by a student during class concerning the relationship between a model and reality.*

In the first part of his statement the student demonstrates a completely distinct differentiation between the model and reality. His epistemological position that a model could be verified however puts into perspective the difference between the world of real appearances and the modeling undertaken by persons. As a consequence of the act of confirmation, which is not further specified, a construct from the world of models is transferred to reality from the point of view of the student.

Together with the implicit premise of the existence of an unquestionable reality, this has dramatic consequences; confirmed models are equivalent to reality, thus they are also equivalent to each other. It is only a small step to the development of hybrid models if the unexplored and often inexperienced application of models in the classroom seem confirmed to the students (in the sense of the above student comments). If model constructs are considered equal to each other, then the differentiation falls into different classes of models as well as the cognitive separation of the individual models. Elements form a certain model can then be incorporated into another model without psychological problems.

Thus a central aspect of the didactic mediation of physical models in class should be the relationship of the models to the observable phenomenon as well as a clear establishment of the limits of the models. If the students cannot distinguish between erwünschten Begriff „Lichtstrahl" argumentiert, hat er die Ebene der realen Erscheinungen noch nicht verlassen, denn er schränkt seine Äußerung durch die Bemerkung, er wisse nicht, ob es ein hundertprozentiger Lichtstrahl sei, stark ein und verweist wiederum auf die starke Bündelung des Laserlichts. Mit dieser Argumentation befindet er sich immer noch auf der Ebene der Realwelt. Diesem Wechsel zurück folgt der Lehrer ebenfalls, indem er zu einer Frage bezüglich des Laserlichts ansetzt. Dies greift der gleiche Schüler wieder durch eine Ebenenvermengung auf: Es handele sich um stark gebündeltes Licht und nicht um einen Lichtstrahl. Da er die Begründung für seine Aussage nicht liefern kann, ist diesem Schüler der von ihm artikulierte Ebenenwechsel allerdings nicht bewusst. Erst ein zweiter Schüler benennt abschließend den Lichtstrahl ausdrücklich als Modellvorstellung, worauf der Lehrer mit dem Unterricht fortfährt.

Die in diesem Unterrichtsausschnitt dargestellten Probleme haben für den Modellgebrauch weitreichende Konsequenzen. So sind Modellmischungen und die Ausbildung von Hybridvorstellungen bei den Schülerinnen und Schülern nicht nur und nicht immer durch einen unreflektierten Umgang mit Modellen zu erklären. Vielmehr können diese Modellmischungen direkt durch den bei vielen Lernenden beobachtete Gleichsetzung von Realität und Modell bzw. der beliebig nahen Annäherung des Modells an die Realität (siehe Abbildung 10) begründet werden.

> **Schüler:** ... Dann schafft man ein Modell, also eine Vorstellung, so könnte es sein. Und das ist meiner Meinung nach ein Modell. Und das ist ja nicht gleichzusetzen mit der Realität, dass es wirklich so ist. Also ein Modell ist ja so lange kein Modell, bis es eindeutig bewiesen ist.

*Abbildung 10: Schüleräußerung im Unterricht zur Beziehung von Modell und Realität*

Durch seine Äußerung zeigt der Schüler im ersten Teil eine ausgesprochen deutliche Differenzierung zwischen Modell und Realität. Seine erkenntnistheoretische Position, dass ein Modell bewiesen werden könne, relativiert diese Unterscheidung zwischen der Welt der realen Erscheinungen und der von Menschen vorgenommenen Modellierungen jedoch. Durch den nicht näher spezifizierten Vorgang des Beweisens wird aus Schülersicht ein Konstrukt aus dem Bereich der Modellwelt in den Bereich der Realwelt überführt.

Zusammen mit der unausgesprochenen Prämisse der Existenz einer eineindeutigen Realität hat dies drastische Folgen: Bewiesene Modelle sind äquivalent zur Realität, also sind sie auch äquivalent zu the models themselves and in relation to reality, then they are not capable of solving problems consistently and physically correct. They indiscriminately jump back and forth between models, descriptions of phenomenon and approaches of explanation without even realizing that they are jumping.

In contrast, students with a distinct meta-conceptual understanding are capable of changing from one pattern of argument of a model to the level of another model, confidently and problem-oriented, name these changes, determine why, and make a clear distinction between reality and the world of models.

### 3. Conclusion

The results of the case study for the utilization of models in the secondary school level II have established that in practice greater emphasis needs to be placed on the various models as it concerns the teaching of optics. The students need to learn how to effectively separate and apply the models.

The individual models are currently introduced one after the other without placing them next to each other. In order to prove the unreliability of the ray model, it is allegedly compared with the wave model. The photon model is introduced after the failure of the wave model in order to explain the photo effect. The photon model has an exclusive character of a particle in this instance. Since one cannot explain interference and diffraction with the particle model (which is generally hardly used to explain other optic phenomenon), the wave-particle dualism of light is thus consequently introduced. This approach causes the development of at least two misconceptions by students. Firstly, they believe that the photon is a quick particle and secondly, they consider the wave-particle dualism as a requirement to apply the specific model, as it is needed.

It is thus recommended when teaching light in level II classes

- to repeat or to newly convey the character and application of models in physics;
- to repeat the previously identified models of light and compare them in their functions;
- to apply the wave model not only to explain diffraction and interference and teach its failure in its ability to explain the photo effect, but also for optic phenomenon like refraction and reflection;
- to discuss why there are a number of models for light and discuss their benefits and drawbacks in illustrating different optic phenomena.

einander. Sollten durch unhinterfragten und oft unbedarften Gebrauch von Modellen im Unterricht diese den Schülern als (im Sinne der obigen Schüleräußerung) bewiesen erscheinen, ist es nur ein kurzer Schritt zur Bildung von Hybridmodellen. Denn gelten Modellkonstrukte als äquivalent zueinander, fällt die Unterscheidung in unterschiedliche Modellklassen und damit die kognitive Trennung der einzelnen Modelle. Elemente aus einem Modell können dann psychologisch problemlos in ein anderes Modell eingebaut werden.

Ein zentraler Punkt bei einer didaktischen Vermittlung physikalischer Modelle sollte deshalb die Beziehung der Modelle zu den beobachtbaren Phänomenen und eine klare Herausarbeitung der Modellgrenzen im Unterricht darstellen. Zeigen Schülerinnen und Schüler keine Trennschärfe zwischen den Modellen untereinander und in Beziehung zur Realwelt, sind sie nicht in der Lage, konsistent und mit physikalischem Sinngehalt Problemstellungen zu bearbeiten. Sie springen zwischen Modellen, Phänomenbeschreibungen und Erklärungsansätzen gleichsam wahllos hin- und her, ohne diese Sprünge zu erkennen.

Im Gegensatz dazu zeigen sich Schülerinnen und Schüler mit ausgeprägtem metakonzeptuellem Verständnis in der Lage, bewusst und problemorientiert von Argumentationsmustern eines Modells in die Ebene eines anderen Modells zu wechseln, diese Wechsel zu benennen, zu begründen und eine klare Unterscheidung zwischen Realwelt und Modellwelt vorzunehmen.

### 3. Schlussfolgerungen

Die Ergebnisse der Fallstudie zum Modellgebrauch in der Sekundarstufe II legen nahe, dass in der Praxis zur Behandlung der Optik größeren Wert auf die unterschiedlichen Modelle gelegt werden muss. Die Schülerinnen und Schüler sollen lernen, die Modelle bewusst zu trennen und zu nutzen.

Zurzeit werden die einzelne Modelle nacheinander eingeführt ohne sie nebeneinander zu stellen. Lediglich das Wellenmodell wird mit dem Strahlenmodell verglichen, um die Unzulässigkeit des Strahlenmodells herauszustellen. Das Versagen des Wellenmodells zur Erklärung des Photoeffekts wird gelehrt und daher das Photonenmodell eingeführt. In diesem Schritt hat das Photonenmodell meist einen ausschließlichen Teilchencharakter. Da mit dem Teilchenmodell (das häufig nicht weiter an anderen optischen Phänomenen angewandt und erklärt wird) aber die Interferenz und Beugung nicht erläuterbar ist, wird auf diesem Wege der Welle-Teilchen-Dualismus des Lichts eingeführt. Diese Vorgehensweise hat die Herausbildung von mindestens zwei Fehlvorstellungen bei den Lernenden zur Folge. Erstens glauben sie, dass das Photon ein schnelles Teilchen ist und zweitens verstehen

The correct application of models in optics does not only present an increased difficulty for students, but this is also the case in other areas of physics and chemistry. Therefore Antje Leisner is completing her dissertation on the topic of long-term learning of and with models in the secondary school grade level I.

In closing it should again be emphasized that there can be immense differences between the physical and didactic focal points of a successful discussion on models. Richard Feynman gave a definite answer concerning the question on the nature of waves or particles [3] "What is light?": "Neither, but a third!" This third, i.e. quantum electrodynamics, as a physical formalism may describe the experimentally apparent phenomena with sufficient precision.

The interpretation of this physical third however even today reveals model-theoretical problems. Roger Penrose writes in the preface to [7]: "**If appraised properly, the particles and waves in the quantum picture prove to be equal,**" whereas Jürgen Ehlers points out in a second preface for the same book [7]: "Meanwhile it has been achieved in (…) quantum mechanics and (…) quantum theory to construct a mathematic formalism that records many experimental facts in a precise manner, which in a certain sense **avoids the previously used pictures of a wave and a particle that only should be applied under specific, mutually exclusive circumstances**."

The range illustrated here between an equalization of concepts of models and their reciprocal exclusion shows that the third suggested by Feynman on the whole needs to be assessed differently in a didactic manner. A third, fourth or additional model of light, as for instance the phasor model should be clearly separated didactically and only be introduced into class after thorough discussion concerning the limits of models.

It is however necessary to practice

> - meta-conceptual ways of thinking
> - using different approaches of models
> - at different levels of the model

during class. The utilization of computer simulation programs that allow a model change during the illustration of simulation processes is thus one of the most promising methods.

**Bibliography**

[1] Johannes Werner: Vom Licht zum Atom. Ein Unterrichtskonzept zur Quantenphysik unter Nutzung des Zeigermodells, Logos Verlag, Berlin 2000.

sie den Welle-Teilchen-Dualismus als Vorschrift, jeweils das Modell zu nutzen, das sich anbietet.

Es empfiehlt sich folglich in der Sekundarstufe II bei der Behandlung des Lichts

- das Wesen von und den Umgang mit Modellen in der Physik zu wiederholen bzw. neu zu vermitteln;
- die bis dahin bekannten Modelle des Lichts zu wiederholen und in ihren Funktionen nebeneinanderzustellen;
- das Wellenmodell nicht nur zur Erklärung der Beugung und Interferenz zu benutzen und sein Versagen bei der Erklärung des Photoeffekts zu lehren, sondern ebenso für optische Phänomene wie Lichtbrechung und Reflexion;
- zu diskutieren, warum es zum Licht mehrere Modelle gibt und deren Vor- und Nachteile beim Darstellen verschiedener optischer Phänomene.

Der richtige Umgang mit Modellen stellt nicht nur in der Optik eine erhöhte Schwierigkeit für die Schülerinnen und Schüler dar, sondern auch in den anderen Gebieten der Physik und Chemie. Daher befasst sich das Promotionsvorhaben von Antje Leisner mit dem langfristigen Lernen von und mit Modellen in der Sekundarstufe I.

Zum Schluss soll hier noch einmal betont werden, dass sich zwischen fachphysikalischen und fachdidaktischen Gesichtspunkten einer erfolgreichen Modelldiskussion zum Teil erhebliche Unterschiede zeigen können. So gab Richard Feynman auf die Frage bezüglich der Wellen- oder Teilchennatur [3] „Was ist Licht?" die eindeutige Antwort: „Keins von beiden, sondern etwas Drittes!" Dieses Dritte, die Quantenelektrodynamik, mag als fachphysikalischer Formalismus die sich experimentell zeigenden Phänomene mit hinreichender Genauigkeit beschreiben.

Die Interpretation dieses fachphysikalisch Dritten offenbart jedoch auch heute noch modelltheoretische Probleme. So schreibt Roger Penrose im Vorwort zu [7]: **„Bei geeigneter Betrachtung erweisen sich Teilchen und Wellen im Quantenbild als gleich."**, während Jürgen Ehlers in einem zweiten Vorwort zum gleichen Buch [7] betont: „Inzwischen ist es zwar gelungen, in der (...) Quantenmechanik und der (...) Quantenfeldtheorie einen viele experimentelle Tatsachen sehr genau erfassenden mathematischen Formalismus zu konstruieren, der in einem gewissen Sinn die früher benutzten **Bilder der Welle und des Teilchens vermeidet und diese als nur unter bestimmten, einander ausschließenden Umständen anwendbar erweist.**"

Die sich hier zeigende Spannweite zwischen einer Gleichsetzung von Modellbegriffen und ihrer ge-genseitigen Ausschließung zeigt, dass das von Feynman angeführte Dritte fachdidaktisch gänzlich anders zu bewerten ist. Ein drittes, viertes oder weiteres Lichtmodell, wie beispielsweise das Zeigermodell, sollte didaktisch klar getrennt und nur unter ausführlicher Erörterung der Modellgrenzen im Unterricht eingeführt werden.

Dabei ist es unerlässlich, dass im Unterricht

- auf verschiedenen Modellebenen
- unter Nutzung verschiedener Modellwege
- metakonzeptuelle Denkweisen

eingeübt werden. Der Einsatz von Computer-Simulationsprogrammen, die einen Modellwechsel bei der Darstellung des Simulationsprozesses zulassen, ist dabei einer der erfolgversprechendsten Ansätze.

**Literatur:** *Siehe linke Spalte.*